\documentclass[aps,prb,twocolumn,showpacs,amsmath,amssymb,superscriptaddress]{revtex4-1}

\usepackage{color}
\usepackage{appendix}

\usepackage{graphicx}
\usepackage{dcolumn}
\usepackage{amsmath}
\usepackage{amssymb}

\usepackage{color}
\usepackage{epstopdf}

\newcommand{\erfc}{\mathrm{erfc}}
\begin{document}
\title{Formation and fragmentation of quantum droplets in a quasi--one dimensional dipolar Bose gas}

\author{S. De Palo}
\affiliation{CNR-IOM-Democritos National 
UDS ViaBonomea 265, I-34136, Trieste, Italy}
\affiliation{Dipartimento di Fisica Teorica, Universit\`a Trieste, Strada Costiera 11, I-34014 Trieste, Italy}
\author{E. Orignac}
\affiliation{Univ Lyon, Ens de Lyon, Univ Claude Bernard, CNRS, Laboratoire de Physique, F-69342 Lyon, France}
\author{R. Citro}
\affiliation{Dipartimento di Fisica ``E. R. Caianiello'', Universit\`a
  degli Studi di Salerno and CNR-SPIN, Via Giovanni Paolo II, I-84084 Fisciano (Sa), Italy}

\begin{abstract}

We theoretically investigate the droplets formation in a tightly trapped one-dimensional dipolar gas of bosonic atoms. When the strength of the dipolar interaction becomes sufficiently attractive compared to the contact one, we show how a solitonic-like density profile evolves into a liquid-like droplet on increasing the number of particles in the trap. The incipient gas-liquid transition is also signalled by a steep increase of the breathing mode and a change in sign of the chemical potential. Upon a sudden release of the trap, varying the number of trapped atoms and the scattering length, the numerical solution of a time-dependent generalized Gross-Pitaevskii equation shows either an evaporation of the cloud, the formation of a single self-bound droplet or a fragmentation in multiple droplets.
These results can be probed with lanthanides atoms and help in characterizing the effect of the dipolar interaction in a quasi-one-dimensional geometry .

\end{abstract}
\date{\today}
\maketitle
\section{Introduction}
\label{sec:intro}

The competition of short and long-range interactions in condensed matter systems gives rise to a rich variety of phases ranging from stripes\cite{loew1994,seibold1998} or checkerboard phases\cite{vanloon2015} to a periodic lattices in two dimensions\cite{PhysRevB.73.014524}, or to  droplet phases in three, two and quasi-one dimension\cite{karpov2021}. The advent of Bose-Einstein condensation of lanthanides atoms\cite{lu2010,Lev_2011,Ferlaino_2012} with large magnetic moments (like $^{164}\mathrm{Dy}$ or $^{166}\mathrm{Er}$) has opened the fascinating perspective of long range dipolar interactions in bosonic systems \cite{chomaz2022dipolar}. Recently, self-bound droplets\cite{luo2021,boettcher2021} have
been unexpectedly observed\cite{Pfau_prl_2016}, associated to a roton minimum of the
excitation spectrum\cite{Chomaz2018}, leading to the subsequent realization of dipolar
supersolids in cigar shaped potentials\cite{Ferlaino_science_2016,Modugno_2019,Pfau_2016,Santos_2016}. In fact, it has been shown, both theoretically and experimentally, that increasing the harmonic confinement along the direction of the dipoles, the ground state of the system passes from a single droplet to multiple-droplet phase\cite{Pfau_prl_2016,Ferlaino_2016}. Quantum droplets have also been very recently observed in a two dimensional dipolar gas in the presence of an isotropic optical dipole trap\cite{ferlaino_nature_2021} confirming the realization of a supersolid phase. Beyond dipolar gases, droplets have also been observed in attractive bosonic mixtures\cite{tarruell_science_2018,Semeghini_prl_2018}. In these systems, spherical droplets form due to the balance (at the mean-field level) close to the collapse threshold of competing attractive and repulsive forces, making first-order Lee-Huang-Yang correction due to quantum fluctuations relevant\cite{Petrov_2015}.
\\
When dimensionality is reduced, quantum fluctuations are enhanced and compete with interactions to determine the nature of the ground state and excitations. To study the interplay between the effective one-dimensional dipolar interaction in a tight transverse harmonic potential, and quantum fluctuations due to reduced dimensionality, we combine a variational approximation of the ground state energy\cite{de_palo_variational_2020} of a uniform dipolar gas with generalized Gross–Pitaevskii equation (GGPE).\\
It has been previously shown
that, in the case of attractive dipolar interaction, an instability appears in  at a critical density: a deep minimum and the steep enhancement of the breathing mode were signals of an incipient instability\cite{depalo_2021}.
The nature of this instability is here analyzed by looking at the evolution of both the chemical potential and the density near the center of the trap. It is found that the density profile evolves from a bright soliton shape\cite{boettcher2021} into a droplet with a flat top shape when the chemical potential turns from positive to negative values at increasing atom number.
Varying the interaction strength, the soliton and droplet regimes can be smoothly connected. We numerically determine their density profile for a broad range of atom numbers and interaction strengths, and map out the boundary of the region that separates bright solitons from quantum droplets. Beyond statics, the time evolution of the droplet when the trap is suddenly released is analyzed. The droplet can either remain in a self-bound state with slight oscillations in time, or fragment for a sufficiently high number of particles. Generally, the fragmentation of a droplet takes place when density oscillations establish at its edge and amplify until they break the droplet. The fragmentation process can be understood using an Euler-like equation, showing that under the effect of negative compressibility, low density regions get depleted of atoms to the benefit of high density regions, leading to a final state formed of dense stable clouds separated by the vacuum.
Our analysis shows that even when the long range tail of dipolar
attraction stabilizes  a  quantum
droplet, the excess density at the center
resulting from longitudinal confinement can result in fragmentation upon a sudden release of the trap. \\

The paper is organized as follows: In Sec.\ref{sec:model} we give the model and derive the generalized Gross-Pitaevski equation (GGPE); in Sec.\ref{sec:Static} we discuss the formation of the droplets when the dipolar interaction is attractive by following the evolution of the chemical potential and density profile for increasing  number of particles; in Sec. \ref{sec:dynamics}, using a time-dependent version of the GGPE, we follow the evolution of the droplet after a sudden release of the trap, showing  a peculiar fragmentation upon increasing the number of particles, N. At varying the contact interaction and N, we follow the evolution from self-bound droplets to fragmented ones and derive a phase diagram. In Sec.\ref{sec:breathing}, we discuss the breathing mode frequency as a marker of droplet formation. Finally, we draw our conclusions in Sec. \ref{sec:conclusions}.

\section{The model and the generalized Gross-Pitaevskii equation}
\label{sec:model}

We consider a system of dipolar bosons in highly elongated traps, {\it i.e.} with a  transverse confining harmonic oscillator frequency
$\omega_\perp $ sufficiently large compared with the longitudinal trapping frequency  $\omega_{ho}$ that it is possible to resort to the so-called single-mode approximation~\cite{Reimann,santos_cond_07} (SMA). The transverse degrees of freedom remain in their ground state, and the effective Hamiltonian for the atomic motion in the longitudinal direction $z$, reads
\begin{eqnarray}
\nonumber
H_{1D}&=&-\frac{\hbar^2}{2 m} \sum_i \frac{\partial^2}{\partial z_j^2}
+g_{1D}\sum_{i<j} \delta(z_i-z_j)\\
&+&\sum_i V_{ext}(z_i) +\sum_{i<j} V_{Q1D}(z_i-z_j),
\label{eq:orig_Ham}
\end{eqnarray}
where $m$ is the mass particle, $V_{ext}(z)= \frac{1}{2}m \omega^2_{ho} z^2$ is the potential energy of the parabolic trap
along the longitudinal $z$-direction,  $g_{1D}$ is the contact interaction coming from Van der Waals or other short-ranged interactions.
Within the SMA the effective 1D dipole-dipole interaction $V_{Q1D}(z)$ ~\cite{Reimann} is
\begin{equation}
\label{eq:vq1d}
V_{Q1D}(z/l_\perp)=V(\theta)
\left[ V^{1D}_{DDI} \left(\frac{z}{l_\perp}\right)
-\frac{8}{3}\delta \left(\frac{z}{l_\perp}\right) \right],
\end{equation}
where
\begin{equation}
\label{eq:v_theta}
V(\theta)=\frac{\mu_0\mu^2_D}{4 \pi}\frac{1-3 \cos^2{\theta}}{4 l^3_\perp}
\end{equation}
encapsulates the sign and the effective strength of the interaction
driven by the vacuum magnetic permeability $\mu_0$, the magnetic dipolar moment $\mu_D$ of the given atomic species, $\theta$ the angle between the dipoles orientation and the longitudinal $z$-axis, and the transverse oscillator length  $l_\perp=\sqrt{\hbar/(m \omega_\perp)}$. In the${}^{162}$Dy case relevant to current experiments~\cite{Lev2020}, $\mu_D=9.93\mu_B$~\cite{tang2018}.
The dimensionless form of effective 1D dipolar potential $V^{1D}_{DDI}$ is :

\begin{eqnarray}
\nonumber
&&V^{1D}_{DDI} \left(\frac{z}{l_\perp}\right)=-2\left| \frac{z}{l_\perp}\right|
+\sqrt{2\pi} \left[ 1+ \left(\frac{z}{l_\perp}\right)^2 \right]
\\
&&e^{\frac{z^2}{2l_\perp^2}} \erfc \left[\left| \frac{z}{\sqrt{2} l_\perp} \right| \right].
\label{V1d_dd1}
\end{eqnarray}
This potential remains finite at $z\to 0$ and thus renormalizes\cite{tang2018} the contact interaction term. As a consequence, an approximation to the ground state energy of a dipolar gas in an elongated trap can be obtained using a Lieb-Liniger ground state wavefunction\cite{de_palo_variational_2020} as trial function in a variational ansatz.

Along the lines of Ref.~\onlinecite{depalo_2021}, we study the dipolar Bose gas in a trap using a generalized Gross-Pitaevskii theory~\cite{pitaevskii1961,gross1963,pitaevskii_becbook}, where we replace the mean-field potential energy of the gas in the standard Gross-Pitaevskii equation by the approximate energy density of the uniform quasi one-dimensional dipolar gas
\begin{equation}
\label{eq:eneden}
e(n)=\frac{\hbar^2}{2 m}  n^3 \epsilon(n),
\end{equation}
where $\epsilon(n)$ is obtained using a variational calculation~\cite{de_palo_variational_2020} of the ground state energy. Such approximation is analogous in spirit to the ones used in Refs.~\onlinecite{kolomeisky2000,oldziejewski_strongly_2019,kopycinski2022} to treat the Lieb-Liniger gas. We expect it to be accurate when the variation of the Gross-Pitaevskii order parameter is sufficiently slow in space and time.

Our approximation to the Gross-Pitaevskii energy functional then reads
\begin{equation}
F_{GP}\!=\!\!\int\!\!dz\!\left[\frac{\hbar^2}{2m} \nabla \phi\! \nabla \phi^*
\!+\!(V_{ext}(z)\! -\!\mu)|\phi|^2 \!+ e(|\phi|^2) \right]\!\!,
\label{eq:fggpe}
\end{equation}
yielding the equation of motion\cite{ohberg_dynamical_2002,oldziejewski_strongly_2019} for $\phi(z,\tau)$, $i \hbar\partial_\tau \phi = {\delta F_{GP}}/{\delta \phi^*}$, \emph{i.e.}
\begin{equation}
\label{eq:ggpe}
i \hbar \partial_\tau \phi=\left[ -\frac{\hbar^2 \nabla^2}{2m} +(V_{ext}(z)-\mu)
+ \frac{1}{\phi}\frac{\delta e(|\phi|^2)}{\delta \phi^*} \right]\phi,
\end{equation}
with the wave function normalized to the number $N$ of atoms in the trap, $N=\int dz |\phi(z)|^2 $.
At zero temperature the condensate is well described by a mean-field order parameter, or “wave function,” $\phi(z,\tau)$ and  this defines an atomic density distribution via $ n(z,\tau ) = |\phi (z,\tau )|^2$. Static solutions satisfy the time-independent Gross-Pitaevskii equation (\ref{eq:ggpe}) with $\partial_\tau\phi=0$.
For the sake of definiteness, we will focus our analysis on the case $\theta=0$ where the attractive dipolar interaction is the strongest.

\section{Phase diagram of the quasi-1D dipolar gas}
\label{sec:Static}

By reducing the strength of the repulsive short-range contact interaction, using for instance Feshbach resonances\cite{feshbach_resonance,duine_feshbach_review},
 the attractive dipolar interaction becomes predominant and enables droplet formation, since the long range attractive part of the inter--particle potential binds the particles while collapse is still prevented by the repulsive core \cite{depalo_2021,oldziejewski_strongly_2019,edler2017}.

Within the Variational Bethe ansatz, for the the${}^{162}$Dy with $l_\perp=57.3 \mathrm{nm}$ we find
that for  $|a_{1D}/a_0| > 6000 $, the equation of state develops a very shallow minimum for extremely low density and, then at larger density, a deep one whose depth increases with $|a_{1D}|$ \cite{de_palo_variational_2020}.
Yet, in order to have a self-bound stable liquid droplet, we need the chemical potential of the system $\mu$ to be negative to prevent the system from evaporating in absence of trap. The liquid density is fixed by the relation for pressure $P= -e(n)+n \mu =0$, and $dP/dn $ should be positive to ensure the stability with respect to excitations.
When this situation is realized, a Maxwell construction can be applied to determine the transition to a liquid droplet (see Appendix \ref{sec:maxwell}).
The fulfillment of these conditions depends on the number of particles in the trap and the scattering length.

\begin{figure}[h]
\begin{center}
\includegraphics[width=85.mm]{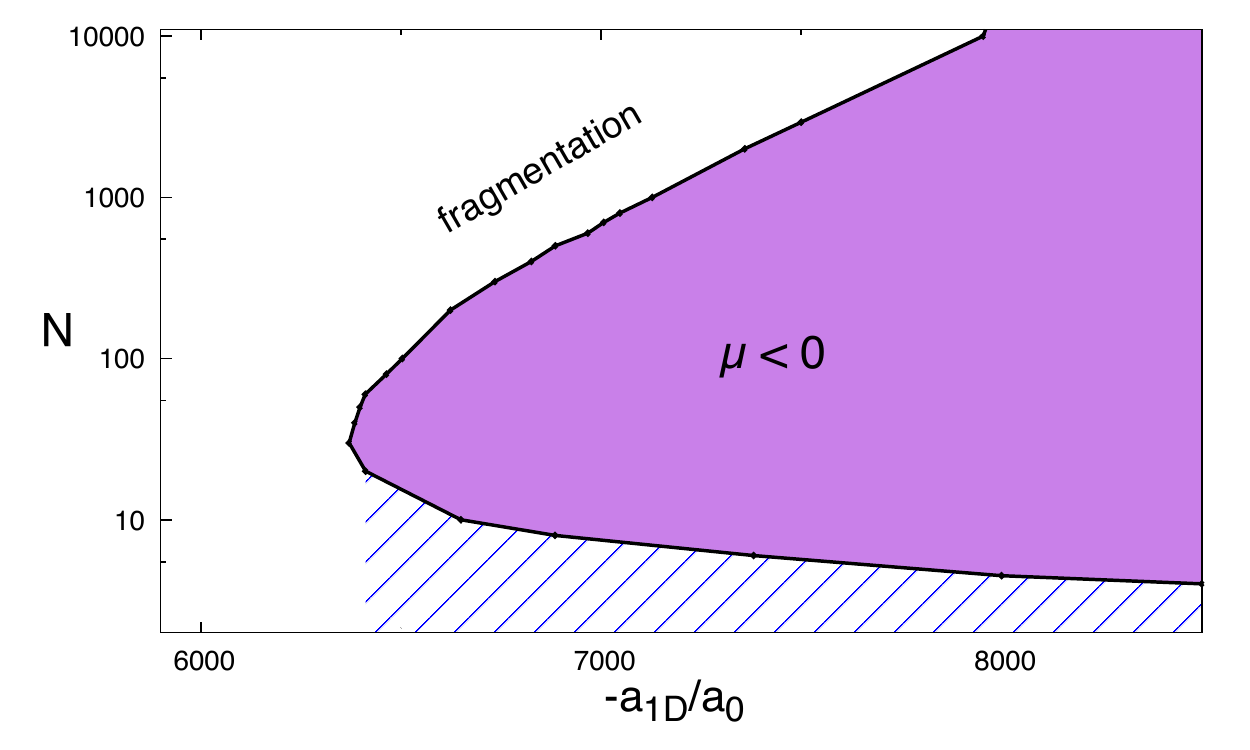}
\end{center}
\caption{Phase diagram for the one-dimensional dipolar boson gas at $\theta=0$ as a function of number of particles in trap and scattering length. In the blue dashed
region, the chemical potential is positive, and density--profiles take a "soliton--like" shape. After releasing the trap, the condensate evaporates. In the violet region, the chemical potential is negative and droplets form in the ground state and they are stable even when the trap is released.  Above the violet region, droplets are still present in the ground state, but upon release of trapping, they fragment into smaller droplets. }
\label{fig:phd_droplets}
\end{figure}

We can construct a phase diagram based on the sign of the chemical potential as shown in Fig.~\ref{fig:phd_droplets},
whose phases we will be detailed in the following.

\begin{figure}[h]
\begin{center}
\includegraphics[width=90.mm]{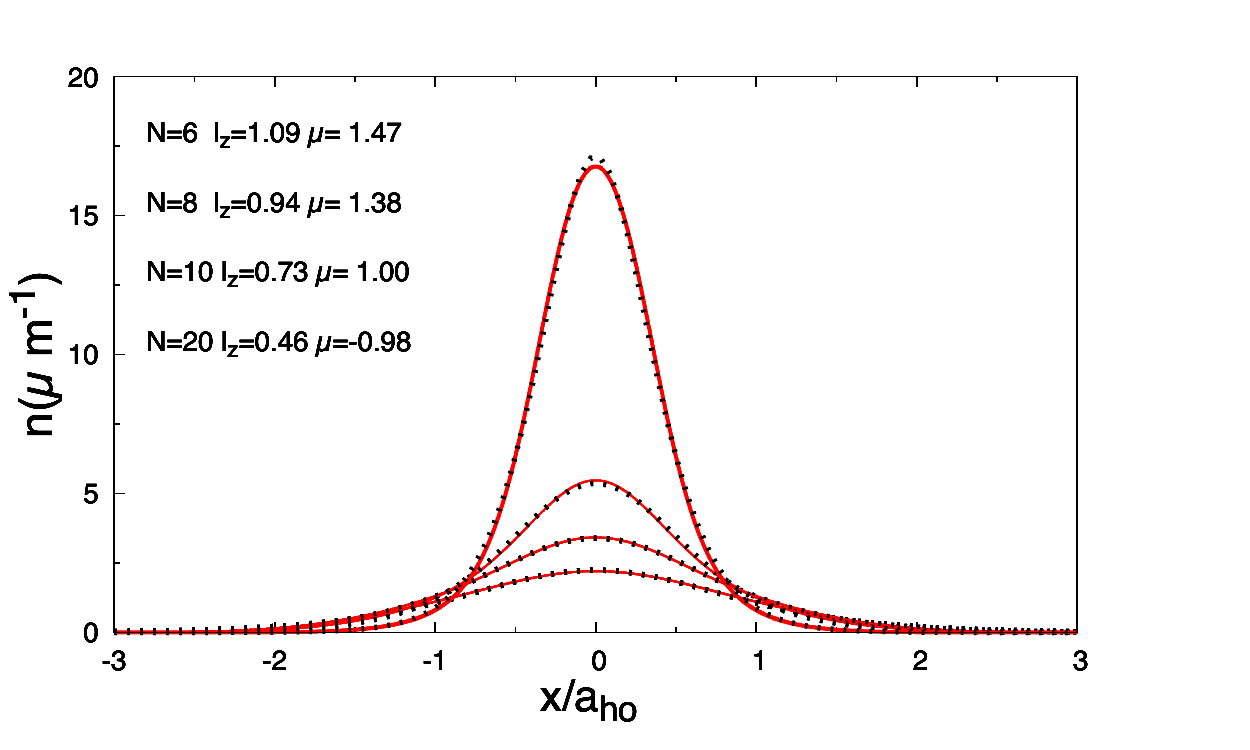}
\end{center}
\caption{Density profiles  for (top to bottom) $N=20,10,8$ and $6$ are shown as solid red curves. Dashed black lines are fits to data using the sech-shape profile $f(x/a_{ho})=a/[e^{x/l_z}+e^{-x/l_z}]^2$. The quality of the fit degrades when the number of particles in the trap is increased.}
\label{fig:soliton_6500}
\end{figure}

We start by considering the region with a small number of particles in the trap ( $N < 10 $, $a_{1D}/a_0 \lesssim -6250 $)
where the chemical potential is positive.
In this region, roughly sketched by the blue dashed area in Fig.~\ref{fig:phd_droplets}, the density profile is well described by solitonic sech-shape. In Fig.~\ref{fig:soliton_6500} for $a_{1d}=-6500 a_0$ we show density profiles for several number of particles and the agreement with the sech-shape profile $f(x/a_{ho})=a/[e^{x/l_z}+e^{-x/l_z}]^2$.
This solution ceases to be a good approximation as soon as the chemical potential becomes negative, which occurs at $N=20$.

\begin{figure}[h]
\begin{center}
\includegraphics[width=90.mm]{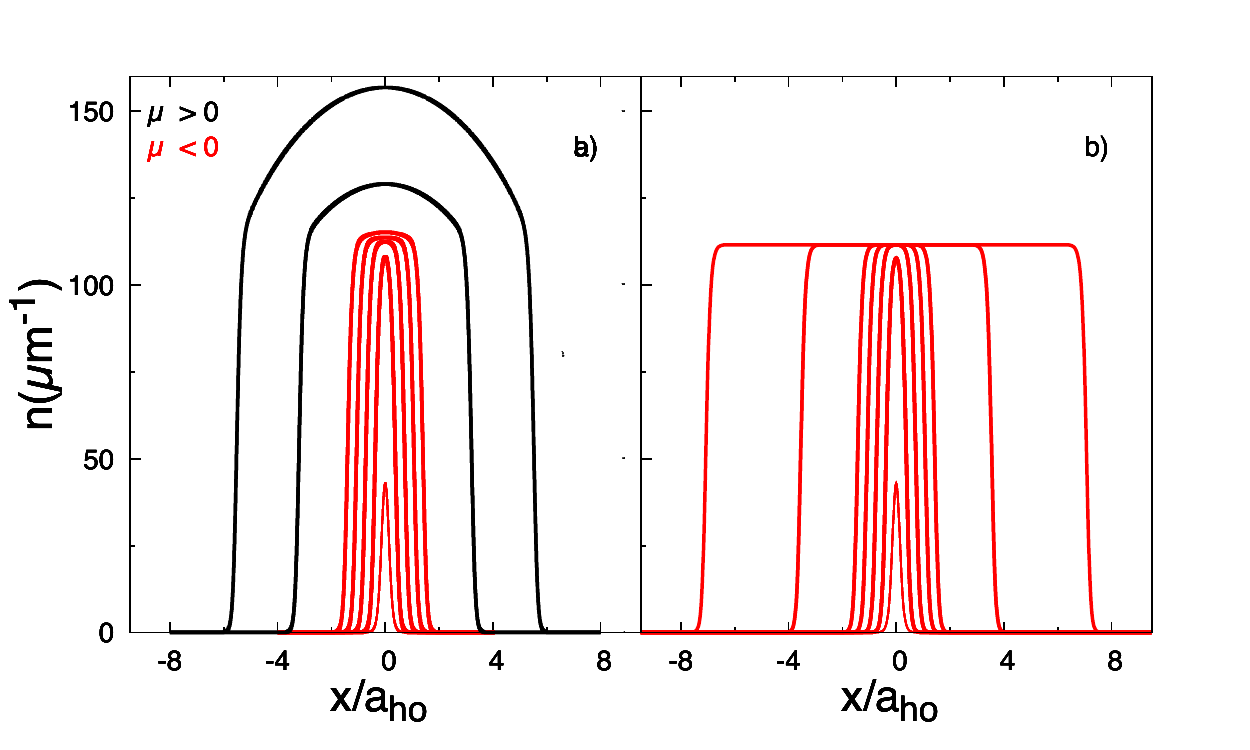}
\end{center}
\caption{ Density profile for the system in the trap are shown in panel $a)$: solid red lines for $N$
for which $\mu < 0$ ($N=20,100,200,300$ and $400$) and black ones for $N=1000$ and $2000$ for which $\mu > 0$ at $a_{1D}/a_0=-7500$. In panel $b)$ we show density profiles for the same number of particle in the absence of
confining potential.}
\label{fig:den_mmu_7500}
\end{figure}

When $\mu$ has become negative the density profile departs from the solitonic shape and slowly approaches the
flat-top shape of a droplet. In panel $a)$ of Fig~\ref{fig:den_mmu_7500} we show density profiles for $a_{1D}/a_0=-7500$
where we can follow the formation of the flat top region at the center
of the trap, together with two density profiles at large $N$ where the chemical potential is again positive
for larger number of particles.
In panel$b)$ we show density profiles for the same number of particles yet in the absence of the the trapping potential
showing the typical droplet shape profiles with the flat top at the center. We note that a Quantum Monte Carlo study of a model with $N=400$ bosons with dipolar attraction and Lennard-Jones repulsion\cite{boninsegni2021} finds free standing one-dimensional droplets for weak enough repulsion.

\begin{figure}[h]
\begin{center}
\includegraphics[width=90.mm]{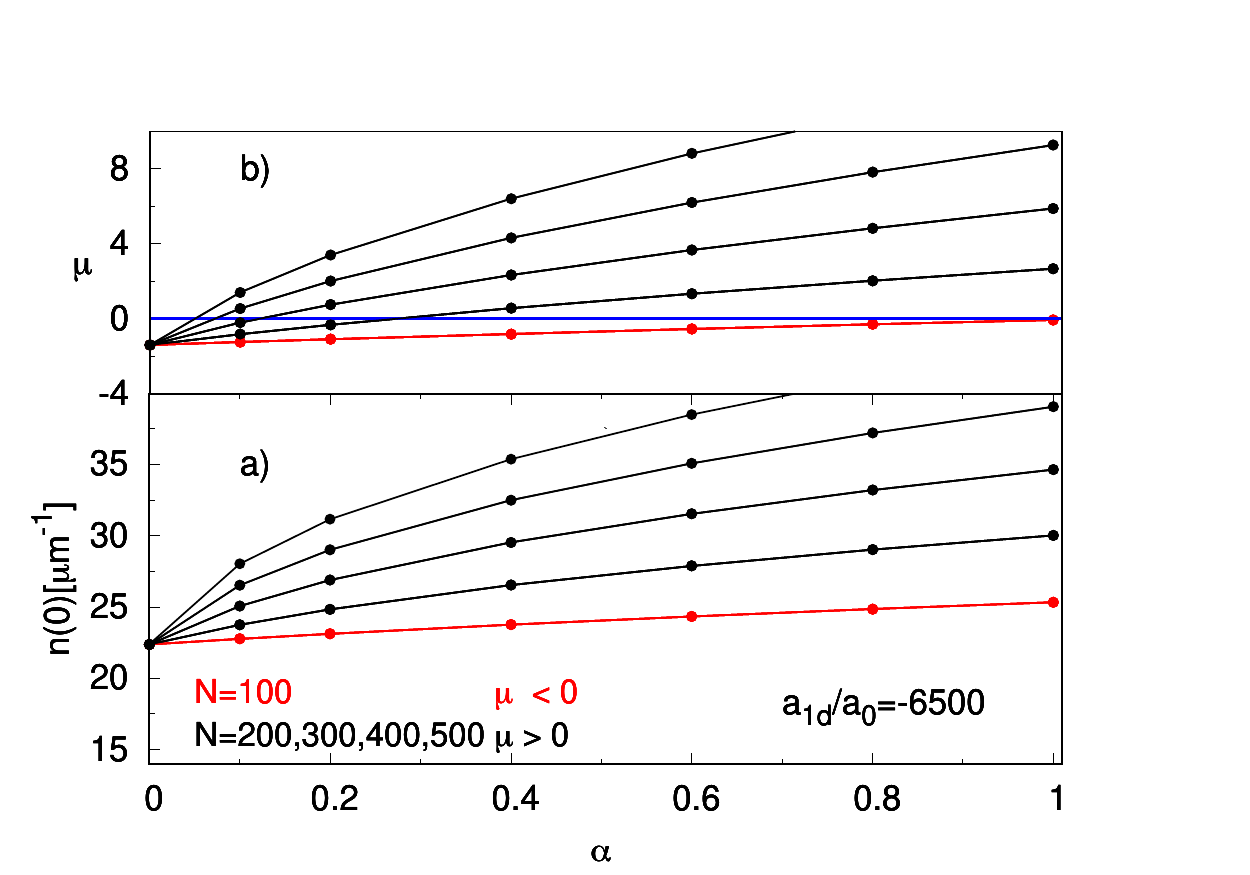}
\end{center}
\caption{ Panel a): density at the center of the trap for atom numbers $N=100,200,300,400$ and $500$ (bottom to top) as a
function of $\alpha (\omega_{ho}^2) $, the strength of the confining potential. In panel b) the chemical potential is shown for the same numbers of atoms as a function of $\alpha (\omega^2_{ho})$. }
\label{fig:scaling_6500_pmu}
\end{figure}

The occurrence of a positive chemical potential for large number of particles turns out to be an effect of the trapping
potential; indeed, in the absence of confinement, the chemical potential is negative and the system shows the
typical density profiles of a droplet. The equilibrium density is reached when the pressure is zero, and when we
increase the number of particles the size of the droplet increases in order to keep the density at the equilibrium
value ( see panel $b$ of Fig.~\ref{fig:den_mmu_7500}).
We assess the effect of the trapping potential by looking at the scaling of the density at center of the trap and of the
chemical potential with the strength of the confining potential (see Fig.~\ref{fig:scaling_6500_pmu}).

\section{Free expansion and fragmentation of droplets}
\label{sec:dynamics}

For the repulsive Lieb-Liniger gas the sudden removal of the trapping potential undermines the stability of the condensate cloud that simply evaporates. However, in the presence of an attractive potential such as the dipolar interaction, the dynamics of the free expansion of the cloud is more varied depending of the number of particles
in the trap and the strength of their interaction.
To study this non-equilibrium problem, we have used a real-time version of the GGPE Eq.~(\ref{eq:ggpe}), with initial condition the ground state wave function in the trap. The latter was obtained using the imaginary-time program to overcome the lesser accuracy of the real-time program.\cite{kumar2015}.

As expected, below a critical number of atoms, in the soliton-like region where $\mu >0$, (blue shaded region in
Fig.~\ref{fig:phd_droplets}) once the confining trap is removed the cloud simply evaporates: the minimum developed in the ground state energy is not sufficiently deep to sustain a self-bound state.

\begin{figure}[h]
\begin{center}
\includegraphics[width=90.mm]{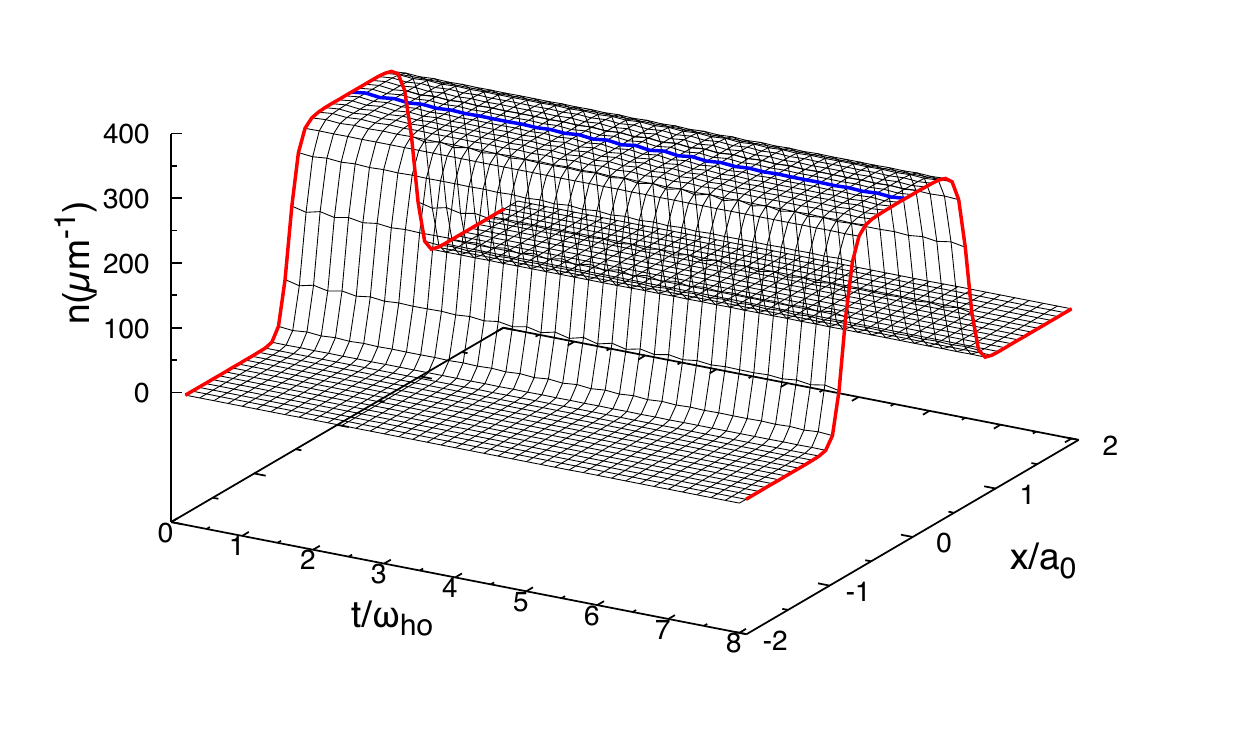}
\end{center}
\caption{Density profile when the trap is suddenly released for $N=600$
and $|a_{1d}/a_0|=8000$ as a function of time. The solid blue line shows
the density oscillation at the center of the trap. The solid red lines show the density profile soon after the release of the trap and for the final observation. }
\label{fig:self_bound}
\end{figure}

When increasing the number of particle in the trap, the chemical potential $\mu$ becomes negative and the density profile tends to the flat-top droplet shape (see the violet region in Fig.~\ref{fig:phd_droplets}
and the red solid lines in panel $a)$ of Fig.~\ref{fig:den_mmu_7500}).
Well inside the violet region of Fig.~\ref{fig:phd_droplets}, where the system is already in flat droplet-shape and the effect of the trap on the density profile is negligible, removing the trap essentially does not alter the droplet profile as shown in Fig.~\ref{fig:self_bound}.

\begin{figure}[h]
\begin{center}
\includegraphics[width=90.mm]{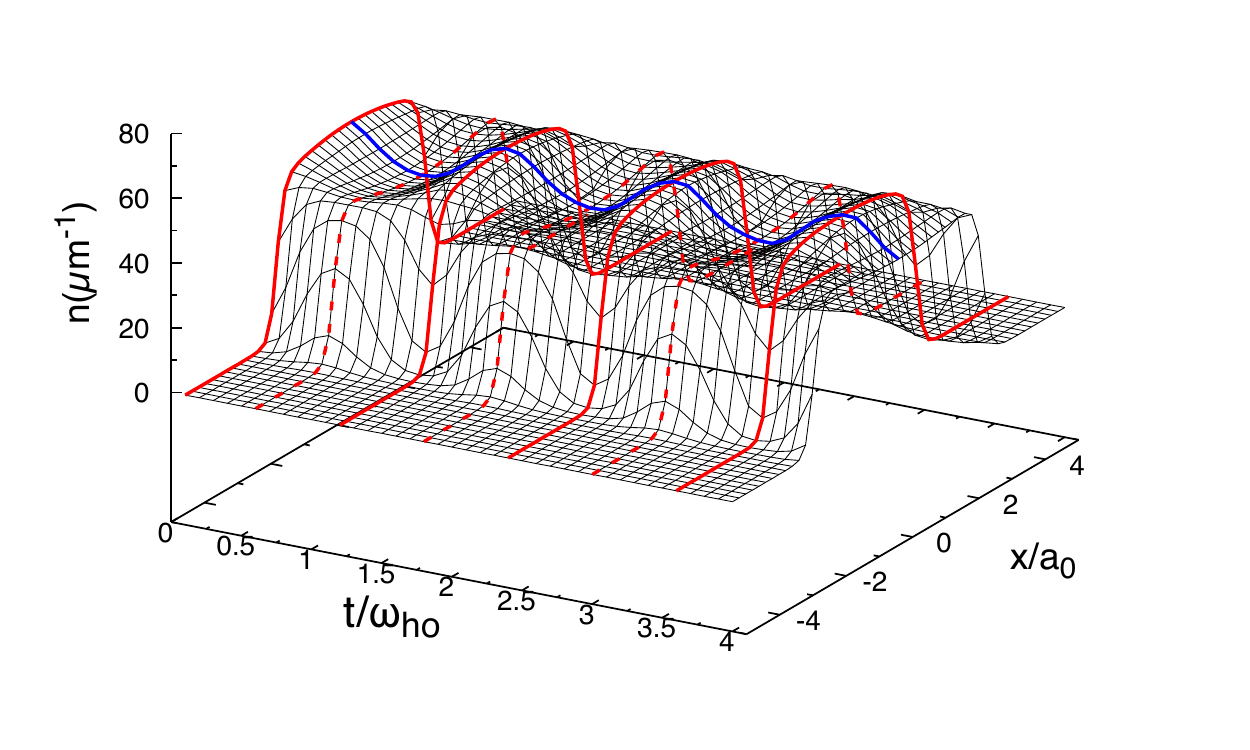}
\end{center}
\caption{Density profile after a sudden release of the trap for $N=300$ and $|a_{1d}/a_0|=7000$ as a function of time. The system oscillates between a regular droplet shape (solid red line) and
another one that show a depression of the density at the center of the profile (red dashed lines). The solid blue line shows the oscillation at the center of the trap.}
\label{fig:density_osci}
\end{figure}

If we move towards the upper boundary in the phase diagram (Fig.~\ref{fig:phd_droplets}), either by increasing the number of particles at fixed scattering length or by reducing scattering length modulus at fixed number of particles, the sudden removal of the
trap leads to an oscillation of the density profile around the equilibrium droplet-shape that preserves the single droplet aspect.
A typical situation is shown in Fig.~\ref{fig:density_osci} where we show the density profile
evolution as a function of time.

\begin{figure}[h]
\begin{center}
\includegraphics[width=90.mm]{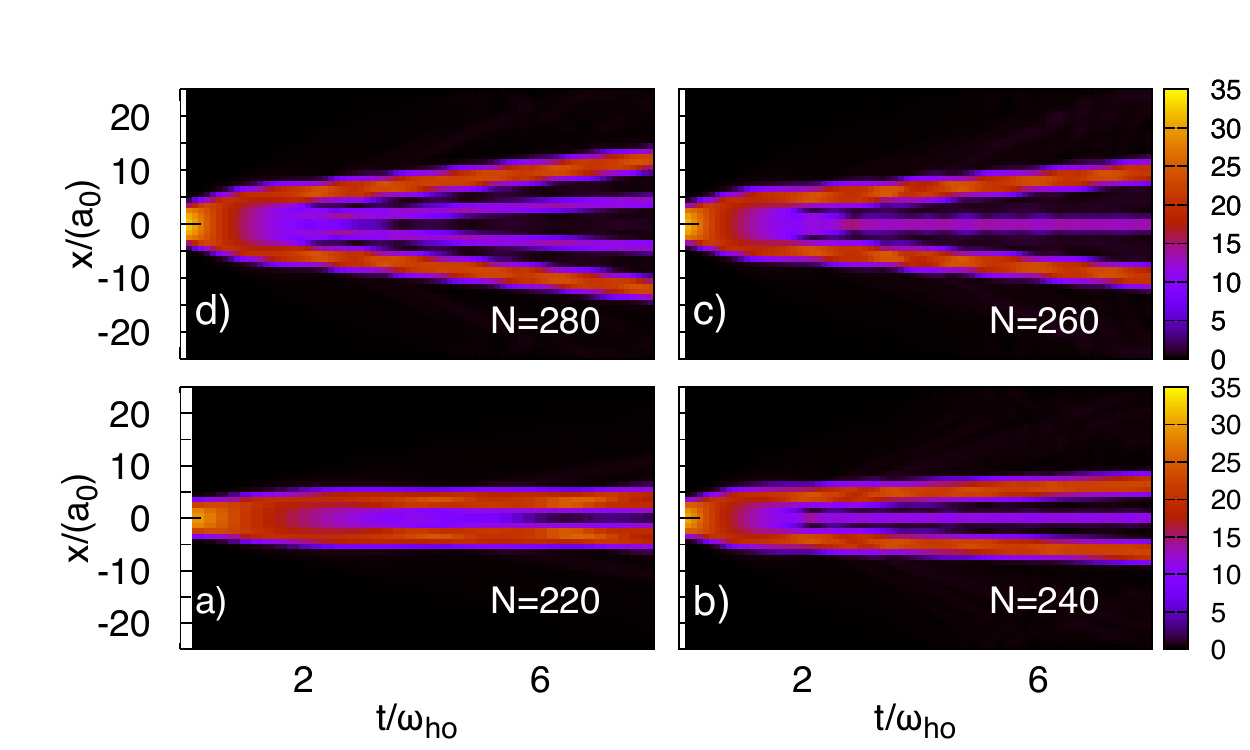}
\end{center}
\caption{Density profile as a function of time after a sudden release of the trap for $a_{1d}/a_0=-6500$ varying the number of particles: $N=220,240,260$ and $280$ in panel
$a)$,$b)$,$c)$ and $d)$ respectively.}
\label{fig:den}
\end{figure}

\begin{figure}[h]
\begin{center}
\includegraphics[width=90.mm]{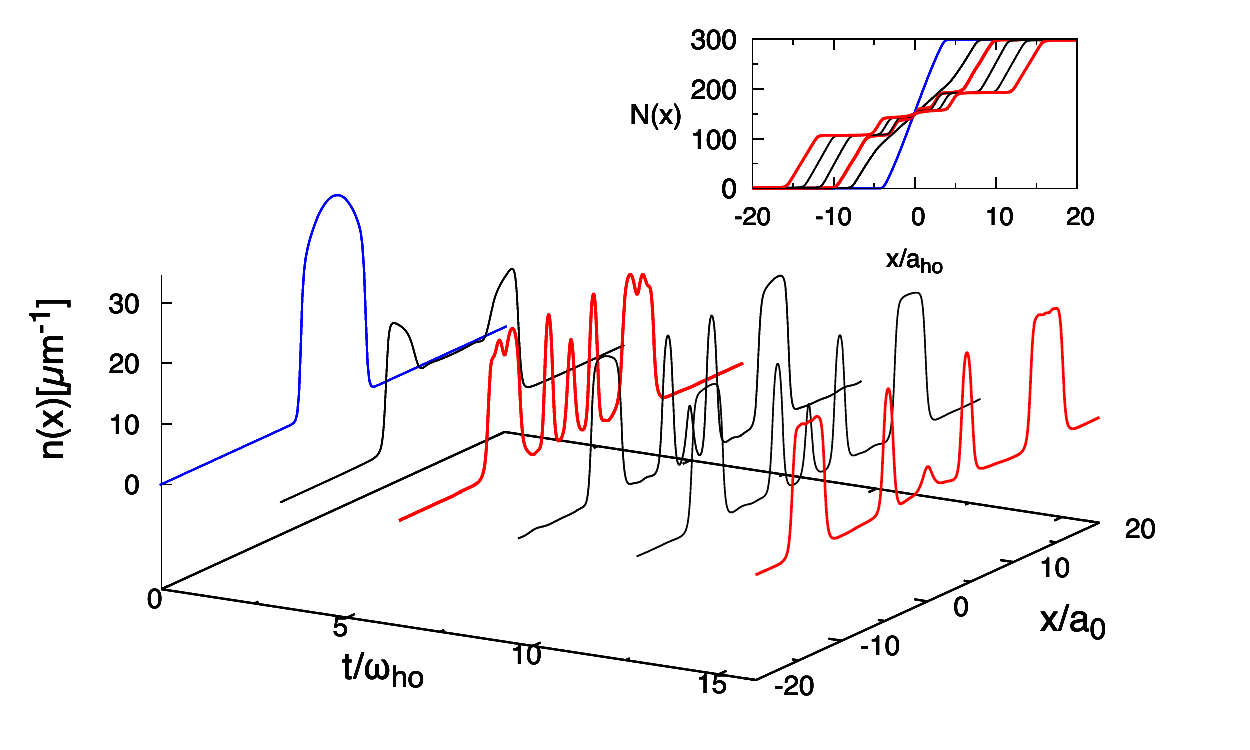}
\end{center}
\caption{In the main panel we show the density profile as a function of time after the the trap is suddenly released
for $a_{1d}/a_0=-6500$ for $N=300$. The blue solid line shows the density profile for $t=0$. The inset provides the quantity $N(x)=\int dx n(x)$ using the same color code of the main panel.}
\label{fig:den_cut}
\end{figure}

Further increasing the number of the particles in the trap,
the chemical potential becomes positive and the sudden release of the trap leads to
the fragmentation of the cloud (see Fig.~\ref{fig:den} and Fig.~\ref{fig:den_cut}) into droplet or soliton-shaped fragments.
In Fig.~\ref{fig:den} we follow the time-evolution of the densities profile, after
the trap has been switched off, for increasing particle numbers, namely $N=220,240,260$
and $280$ at $a_{1D}/a_0=-6500$. The oscillations previously observed (see Fig.~\ref{fig:density_osci}) become so deep
to eventually break the condensate cloud.

In Fig.~\ref{fig:den_cut}, instead, we follow the evolution of the cloud for the case $a_{1D}/a_{ho}=-6500$ and $N=300$, a situation in which
the chemical potential of the system is positive and the density in the trap ( see blue solid line ) shows a build-up of particles in the center induced by the confining potential. Once the trap is released, the potential energy  of these particles converts into kinetic energy, they accumulate at the edges of the cloud and hence form two outer and inner droplets
that steadily move away from the center of the cloud with a constant velocity.
At the center of the trap remains a peak that slowly melts down as its particles escape into the droplets. We can follow the migration of particles from the center of the cloud to the two side-droplets in the inset of Fig.~\ref{fig:den_cut}
where we plot the number of particles $N(x)=\int_{-\infty}^x dx' n(x')$ at the left of position $x$. The outer droplets contain each roughly $\simeq 106$ particles, while the inner ones only $N\simeq 35$, their shape being the one of the systems in the absence of a trap for that number of particles, while the central peak progressively is melting down. \\
Physically, at the initial time the system possesses an excess of potential energy that gets converted into kinetic energy of the particles expelled from the center to form the side droplets moving at constant velocity, as we will show below by using time dependent GGPE equation.

\subsection{Fragmentation of the droplet within the time-dependent GGPE}

In classical physics the fragmentation of a droplet takes place because density oscillations establish at the interface between the liquid and the gas. These can be amplified until they break the droplet. In this process the superficial forces, due to superficial tension, and the inertial forces, due to the relative velocity between the two phases, give rise to the oscillations. Depending on the geometry of the system and the properties of the liquid, some oscillations can propagate amplifying while others can attenuate due to dissipative effects. Those that amplify are causing the fragmentation of the droplet. \\
In our quantum case, we can describe the fragmentation of the droplet within the time dependent GGPE.

Using density and phase variables, $\phi(x,t)=\sqrt{n}(x,t) e^{i\psi(x,t)}$,  we can rewrite the time dependent GGPE in the form
\begin{eqnarray}
  \label{eq:bernouilli}
 && \partial_t \psi +\frac {(\partial_x \psi)^2}{2m} + e'(n) +V(x)\theta(-t) -\frac{1}{2m\sqrt{n}} \partial_x^2(\sqrt{n}) =0, \nonumber\\
  \label{eq:continuity}
 && \partial_t n +\frac{\partial_x (n \partial_x \psi)} m=0.
\end{eqnarray}

\begin{figure}[h]
  \centering
\includegraphics[width=9cm]{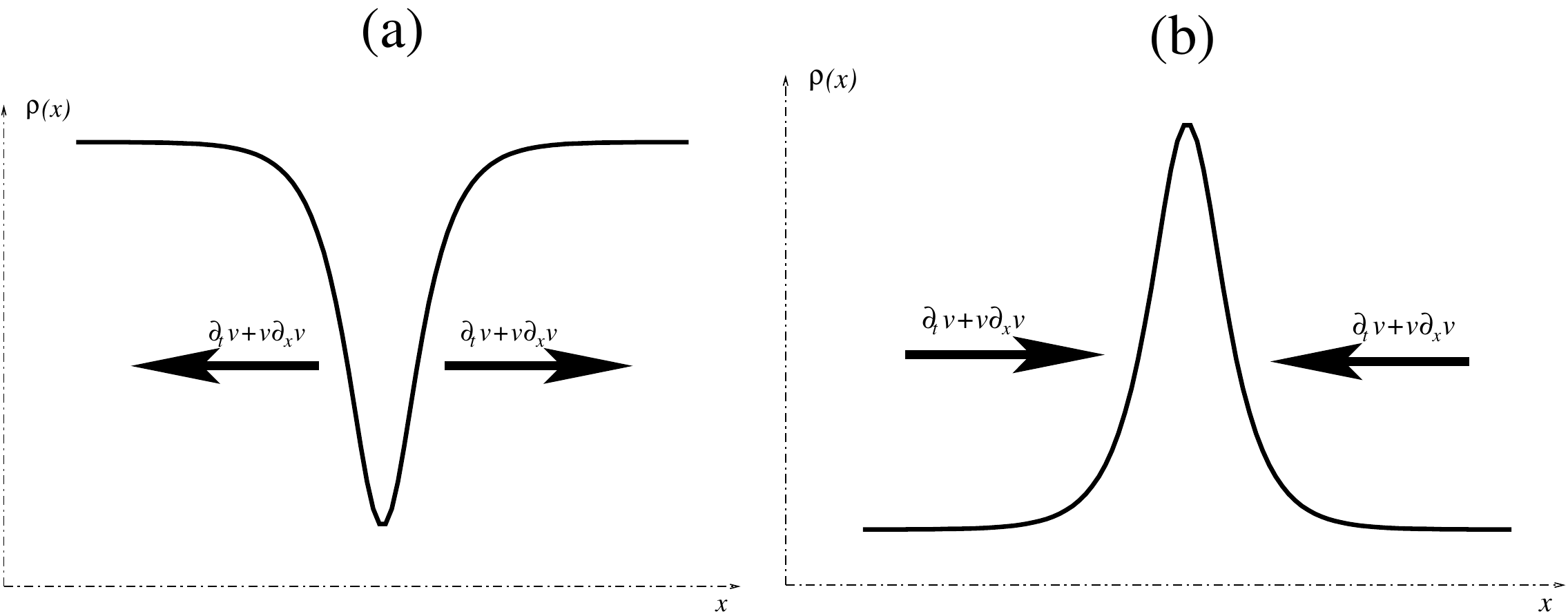}
  \caption{Illustration of the instability when $e"(\rho)<0$. (a) With a minimum of the density, the particles are pushed away from the minimum by the pressure, deepening the minimum with time. (b) With a maximum of the density, the particles are attracted to the maximum, heightening it with time. }
  \label{fig:instablity}
\end{figure}

For $t<0$, we have the static solution with $\partial_t n=0$ and $\partial_t\psi=-\mu$. For $t>0$, the trapping potential is removed. For $t=0$, both $\rho$ and $\psi$ are continuous, therefore, the initial conditions are $\partial_t n(x,0_+)=0$ and $\partial_t\psi(x,0_+)=V(x)-\mu$. After a short time interval $\Delta t$, $\partial_x \psi = \Delta t \partial_x V(x)$, indicating that particles are ejected away from the center of the trap with a velocity gradient.
Differentiating Eq.~(\ref{eq:bernouilli}) with respect to $x$ leads to the Euler equation with a quantum pressure term
\begin{equation}
  \label{eq:euler}
  \partial_t v + v \partial_xv +e"(n)\partial_x n -\partial_x \left[\frac{1}{2m\sqrt{n}} \partial_x^2(\sqrt{n})\right]=0,
\end{equation}
with $v=\partial_x\psi/m$. Let's consider values of $n(x,t)$ such that $e"(n)<0$. If we neglect the quantum pressure term, Eq.~(\ref{eq:euler}) implies that the acceleration $\partial_t v + v\partial_x v$ has the sign of $\partial_x n$. Therefore, (see Fig.~\ref{fig:instablity}) the atoms are pushed away from low density regions and attracted towards high density regions, leading to instability of atomic densities with $e"(n(x,t))<0$. For long enough times, we expect to have only regions where $n(x,t)=0$ or regions with $n(x,t)$ large enough to ensure $e"(n)>0$.
This picture is obtained neglecting completely quantum pressure, an approximation that is justified if the density is sufficiently uniform. Clearly, at the boundary between zero density and stable density regions, the quantum pressure becomes important and ensures a continuous transition between zero and stable density, as explained in App.~\ref{sec:GGPE}. A more precise picture of the effect of quantum pressure can be given at initial times.
Linearizing Eqs. (\ref{eq:bernouilli})-(\ref{eq:continuity}) around the initial equilibrium state, and combining them into a single equation, we obtain
\begin{eqnarray}\label{eq:initial-linearized}
&& \partial_t^2\psi -\left[\frac{e"(n_0)}{m} \frac{\partial n_0}{\partial x} -\frac {\partial_x^2(\ln n_0)}  {4 m^2} \right] \partial_x \psi \\ && -\left[\frac{n_0 e"(n_0)}{m}- \frac {\partial_x^2(\sqrt{n_0})} {m^2 \sqrt{n_0}}\right] \partial_x^2 \psi + \frac{\partial_x n_0}{2m^2 n_0} \partial^3_x\psi + \frac {\partial_x^4 \psi} {4m^2}=0\nonumber
\end{eqnarray}
For long wavelength excitations, we can neglect $,\partial_x^3\psi, \partial_x^4\psi$ in Eq.~(\ref{eq:initial-linearized}). The resulting second order partial differential equation is elliptic\cite{smirnov_pde} when
\begin{equation}\label{eq:ell-cond}
    n_0 e"(n_0) < \frac{\partial^2_x(\sqrt{n_0})}{m\sqrt{n_0}},
\end{equation}
and in such case, disturbances are growing exponentially with time. Using the equilibrium GGPE, the instability condition Eq.~(\ref{eq:ell-cond}) reduces to $n_0 e"(n_0) -2 e'(n_0) +\mu -2V(x) <0$. Since $V(x)$ is maximal and the initial density $n_0$ is lowest near the edges, an instability is favored there.
This leads us to expect at long times the formation of droplets as the condensate expands. In the simplest case of very shallow trapping, the final state can be a single droplet. With a deeper trap, the density near the edges of the trap is low, while the initial acceleration is strong, a situation that favors the breakup of the condensate in multiple droplets at long times. The asymptotic form of the solution can be described as follows.
A droplet moving at uniform velocity is described by a solution $\phi(x,t)=\phi_0(x-vt) e^{i mv x -m \frac{v^2t} 2 -\bar{\mu} t}$, where $\phi_0$ satisfies the static GGPE with effective chemical potential $e(n_1)/n_1<\bar{\mu}<0$.  The final state is made of such moving droplets of increasing velocities as one moves away from the origin. This guarantees that the separation of the droplets increases with time as observed in Fig.\ref{fig:den}. The ansatz for the final state is thus
\begin{eqnarray}
  \phi(x,t)=\sum_j \phi(x-v_j t-x_j,\bar{\mu}_j)e^{i m v_j t -\frac{m v_j^2}2 t -\bar{\mu}_j t}.
\end{eqnarray}
Particle number, momentum, and energy conservation give some constraints on the final state. The energy of the final state is the sum of energies at rest and kinetic energies of the droplets. With two droplets moving at velocities $v$ and $-v$ with $N/2$ particles each,
\begin{eqnarray}
  E_\mathrm{initial}=2E_{\mathrm{droplet}}(N/2) + \frac N 2 m v^2,
\end{eqnarray}
allowing us to find the final velocity of the droplets.

\section{Breathing mode approaching the gas-liquid transition}\label{sec:breathing}

We can see the incoming gas-liquid transition that leads to the formation of the droplets by looking at the breathing mode:
\begin{equation}
\omega^2_b= -2 \langle \sum^N_{i=1} z^2_i\rangle \left[
\frac{\partial \langle \sum^N_{i=1} z^2_i\rangle}{\partial \omega^2_{ho}}
\right]^{-1}
\end{equation}
that strongly departs from the effective Lieb-Liniger behavior\cite{depalo_2021} in close agreement with the the recent experiments \cite{Lev2020}. Approaching an incompressible liquid is signalled by the increase  of the breathing mode frequency.

\begin{figure}[h]
\begin{center}
\includegraphics[width=90.mm]{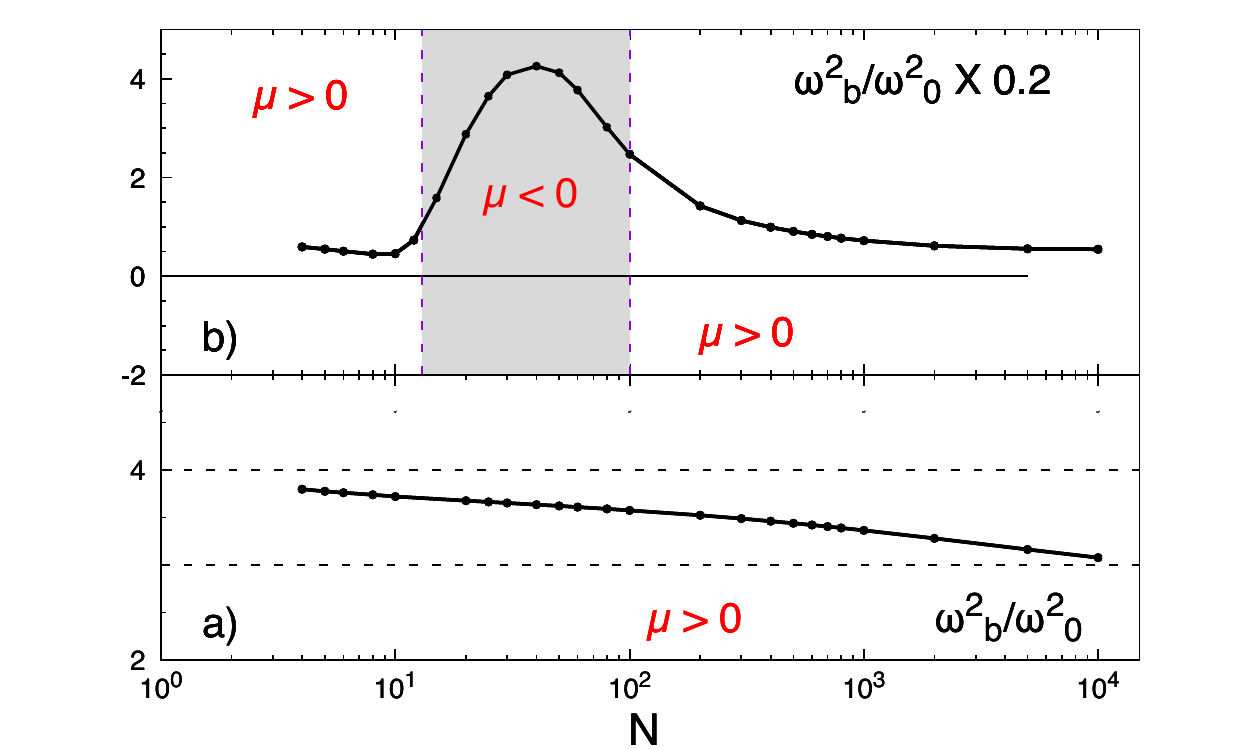}
\end{center}
\caption{Breathing mode frequency (solid black dots)
is shown for the case $\theta=0$ as function of the number of particles in the trap for fixed $a_{1d}/a_{0}=-1000$ and
$-6500$  in panel $a) $ and $b) $ respectively.
In panel $b)$ the breathing mode frequency  has been reduced by a factor $1/5$. In the gray shaded area of panel $a)$ the chemical potential of the system is negative.}
\label{fig:chem_a1d_6500}
\end{figure}

In Fig.~\ref{fig:chem_a1d_6500} we show the breathing mode frequency, computed along the lines of Ref.~\onlinecite{depalo_2021} as a function of the particle number $N$ inside the trap for two selected coupling strengths: $a_{1d}=-1000 a_0$ and $-6500 a_0$ respectively in panel $a)$ and $b)$. When the strength of the  dipolar component is small compared to the contact interaction one, so that, despite its different sign, the system follows the Lieb-Liniger physics once we have "renormalized" the contact strength\cite{depalo_2021} with an effective strength that takes care of the short-range
dipolar interaction. The breathing mode $\frac{\omega^2_B}{\omega^2_{ho}}$ is indeed
a monotonic function bounded between $4$ (the low-density limit) and $3$, the chemical potential is always positive (see panel $a)$ of Fig.~\ref{fig:chem_a1d_6500}) and the condensate needs the external confinement to form a cloud.

This behavior changes when the role of the attractive dipolar interaction becomes predominant and a minimum occurs in  the equation of state whose depth gets larger on increasing  $|a_{1d}|$. When the number of particles in the trap is small,
the chemical potential is positive and breathing mode decreases on increasing $N$, as the in the "renormalized" Lieb-Liniger behaviour.
On increasing the number of particles the chemical potential becomes
negative followed by a steep increase of the breathing mode frequency as a function of the number of
particles in the trap. The values in this region largely above the Lieb-Liniger upper bound and eventually diverge for very large $|a_{1D}|$, denoting the increasing rigidity of the system.

\section{Discussion and conclusions}
\label{sec:conclusions}
In conclusion, we have studied the droplets formation and fragmentation in a trapped quasi-one dimensional dipolar Bose gas. This system is described within an effective Lieb-Liniger gas model whose ground state energy is determined by a variational Bethe-ansatz approach. In the case of attractive dipolar interaction ($\theta=0$) we have shown, using a generalized Gross-Pitaevski equation, that the instability in the system at increasing the scattering length is associated with the formation of a droplet. The transition is signalled by the sign change of the chemical potential, as well as the appearance of a sufficiently deep minimum in the bulk energy as a
function of density. On increasing the number of particles the density profile shows a soliton-like shape, typical of a trapped Lieb-Liniger gas, that becomes a flat-top when the droplet is formed. The scaling of the density at the center of the trap at increasing number of particles is an other indicator of the droplet phase. The density doesn't scale with number of particles N deep in the droplet phase, and scales linearly with N away from it.
Analogous results were obtained in Ref.\onlinecite{baizakov2009} with hard core bosons and in Ref.~\onlinecite{oldziejewski_strongly_2019}, with Lieb-Liniger bosons when the dipolar attraction was added to as a perturbation the generalized Gross-Pitaevskii Equation\cite{kolomeisky2000,ohberg_dynamical_2002}.
Using a more accurate expression, we have confirmed the presence of solitons and droplets stabilized by dipolar attraction. Furthermore, we have also analyzed the evolution of the droplet after a sudden release of the trap, by using a real-time dependent version of the GGPE. The evolution of the droplet shows a rich scenario depending on the number of particles and the scattering length. At higher scattering length $|a_{1D}|$ the droplet remains self-bound oscillating around the center of the trap or undergoes a blocking phenomenon at even larger scattering length. For intermediate values of the scattering length, and few hundreeds of particles, the droplet fragments and the density passes from the center of the trap towards the edges of the trap where two droplets are formed. Finally for very low scattering length the droplet simply melts at longer times. The fragmentation of a droplet after a sudden trap release is a phenomenon that could be observed in lanthanide atoms for trapped in quasi-one-dimensional geometry. \\
The experimental system  currently under investigation is made of an array of decoupled tubes\cite{tang2018,Lev2020},
each containing a certain number of particles $N_i$, where the tubes with the largest number of particles are at the center while the outer tubes contain a smaller number of particles. Since the gas-liquid transition depends on the number of particles in the tube, at fixed scattering length, in principle one can perform the following cycle: First switch off the longitudinal trapping and let the clouds expand in 1D. The condensates in tubes  where $N_i$ is small will melt and fragment in tubes where $N_i$ is too large, while in some tubes there could be just enough particles
 to form a single self-bound state. In such a situation, the clouds with inadequate $N_i$  evaporate or fragment, while the ones forming droplets stay in place. Then if the system is confined again, the atoms in the self-bound states and a fraction of atoms in the tubes with condensate fragmentation will remain.  After a few cycles, only the self-bound droplets at lower temperatures would remain.
 Let us finally stress that our results have been obtained in the case of very tight transverse trapping so that the single mode approximation is applicable. In the case of a less tight transverse trapping, some energy could be gained by allowing the transverse wavefunction to be more spread out than in the SMA approximation. Such an effect could be especially important in the case of a tighter longitudinal trapping that gives rise to an excess density compared with the regular droplet. The investigation using a more general variational ansatz taking into allowing variation of the transverse width of the trapped gas\cite{salasnich2004,salasnich2005} is under way.
\begin{acknowledgments}
We thank M. Salerno and B. Lev for enlightening discussions.
\end{acknowledgments}
\appendix

\section{The Maxwell construction}
\label{sec:maxwell}

Let's consider a system whose ground state energy per unit length $e_0(n)$ has a local maximum at $n=0$, a single minimum at $n=n^*>0$, and grows to infinity with $n$.
These conditions imply that  $\frac{d^2e_0}{dn^2}$ changes sign at $n=n_i$ with $0<n_i<n^*$.
For $0<n<n_i$, $\frac{d^2e_0}{dn_i^2}<0$, so the system has a negative compressibility and becomes unstable against phase separation.
The stable ground state of a system of length $L$, at a nominal density $n=N/L$, is formed of a  stable region of length $x$, with a density of atoms $n_1>n_i$, that we
call the droplet, surrounded by a vacuum of size $L-x$.
The conservation of atom number yields $N=n L= n_1 x$ \textit{i. e.}
$x/L=n/n_1$.  The actual ground state energy is
\begin{equation}
  \label{eq:energy-ps}
  E(n,n_1)=e_0(n_1) x + e_0(0) (L-x),
\end{equation}
giving a ground state energy per unit length
\begin{equation}
  \label{eq:en-gs-perlength}
  e(n,n_1)=e_0(n_1)n/n_1+e_0(0)(1-n/n_1).
\end{equation}
The density $n_1$ is obtained by minimizing energy $e(n,n_1)$ at fixed $n$, giving
\begin{equation}
  \label{eq:min-n1}
  \frac{de_0}{dn_1}=\frac{e_0(n_1)-e_0(0)}{n_1}.
\end{equation}
Geometrically, $e(n,n_1)$ represents a line segment of the secant to
the graph of $e_0(n)$ joining the points $n=0$ and $n=n_1$, and Eq.~(\ref{eq:min-n1})
indicates that we minimize the energy by choosing the tangent to the
curve $e_0(0)$ that goes through the origin\cite{landau-statmech-english}. This guarantees that the function $e(n)$ is
less than $e_0(n)$ for $0<n<n_1$.  To summarize, the Maxwell
construction shows that with $n_1$ determined by (\ref{eq:min-n1}) ,
for $n<n_1$ our system will form a droplet of density $n_1$ and size
$x=L n/n_1$ with ground state energy given by
Eq.~(\ref{eq:en-gs-perlength}).  For densities greater than $n_1$ the
homogeneous system is stable. So concluding, $n_1$ is the smallest possible ground state density for a homogeneous system.

Now, instead of considering a fixed number of particles, let's consider a
fixed chemical potential. We work with the pressure\cite{landau-statmech-english},
$p=e_0(n)-\mu n$. According to the previous results,
with $\mu \le e(n_1)/n_1$ the ground state energy is minimal
when $n=0$. With $\mu> e(n_1)/n_1$, the minimum of the
ground state energy energy is for $n > n_1$. Therefore, at $\mu=e(n_1)/n_1$, the density
jumps from $n=0$ to $n=n_1$. Graphically, $n_1$ is determined from
\begin{equation}
\int_0^{n_1} dn \left(\frac{de_0}{dn}-\mu \right)  =0
\end{equation}
i. e. the area delimited by $0<n<n_1$ between the graph of $de_0/dn$ and the straight line $de_0/dn=\mu$ is algebraically zero.\cite{landau-statmech-english} The pressure $p$ is vanishing for $0<n<n_1$, that is in the entire droplet phase.

\begin{figure}[h]
\begin{center}
\includegraphics[width=85.mm]{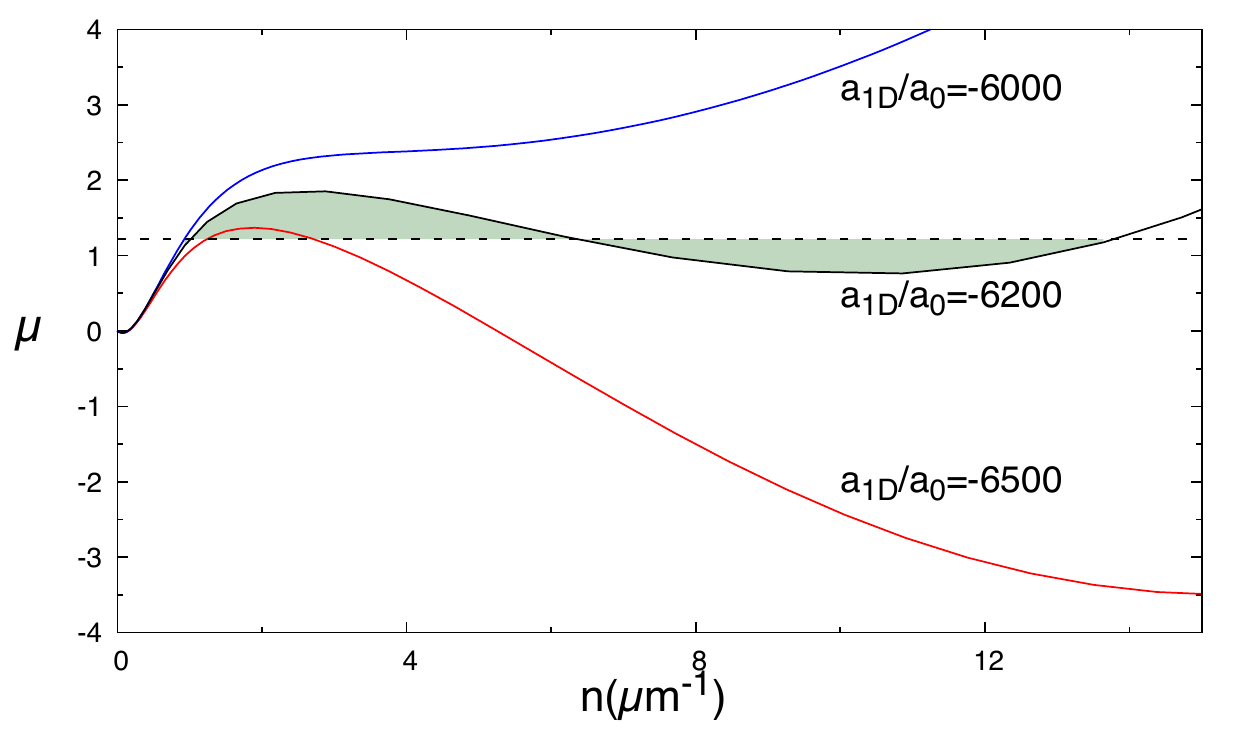}
\end{center}
\caption{Chemical potential from the equation of state computed using
the Variational Bethe Ansatz for three different scattering lengths: namely $a_{1D}/a_{ho}=-6000,-6200$
and $-6500$ from top to bottom. The Maxwell construction, in which the two green-shaded areas are equal, only applies in the $a_{1D}/a_{ho}=-6200$ case among the ones shown.}
\label{fig:chem_a1d}
\end{figure}
The gas--liquid transition can be followed looking at the chemical potential as a function of the particle density as shown in Fig.~\ref{fig:chem_a1d}, where $\mu(n)$ is shown for three selected scattering lengths: namely $a_{1D}/a_0=-6000,-6200$ and $-6500$. At $a_{1D}/a_0=-6000$ there is the hint of the minimum that forms at large density as it will occur at larger $|a_{1D}/a_0|$.

If we take a system in the canonical ensemble in the presence of a potential $V(x)$, the situation is analogous to the case of the grand canonical ensemble. To find the particle density, we must minimize:
\begin{equation}
  \int dx [\epsilon(n(x)) + n(x) V(x)],
\end{equation}
where $\epsilon(x)=e(0) (1-n(x)/n_1)+e(n_1) n(x)/n_1$ when $n(x) \le n_1$ and $\epsilon(n(x))=e_0(n(x))$ when $n(x)\ge n_1$. This yields the condition
\begin{equation}\label{eq:dens-pot}
  \frac {de_0}{dn}=-V(x),
\end{equation}
when $n(x)>n_1$ and
\begin{equation}\label{eq:vac-pot}
  \frac{e_0(n_1)-e_0(0)}{n_1}=-V(x)
\end{equation}
if $n(x)<n_1$.  While the first condition~(\ref{eq:dens-pot}) can be satisfied by varying the density $n(x)$ the second condition~(\ref{eq:vac-pot}) cannot. Physically, this indicates that the system separates into regions with density $n(x)\ge n_1$ with a local density determined by the applied external potential, and vacuum regions. The Eq.~(\ref{eq:vac-pot})  defines the boundaries between these regions.

\section{Generalized Gross-Pitaevskii equation}
\label{sec:GGPE}

Within the Maxwell construction,  we have completely neglected surface tension effects, and allowed for discontinuity of the density in the presence of an external potential.
Now, we return to the generalized Gross-Pitaevskii equation~(\ref{eq:fggpe}) assuming that $e(|\phi|^2)$ satisfies the condition for droplet formation of the Maxwell construction.

From our previous considerations on the grand canonical case, we expect that the densities $n=0$ and $n=n_1$ can coexist when the chemical potential is
\begin{equation}
  \label{eq:mu-crit}
  \mu=\frac{e(n_1)}{n_1}=\left(\frac{de}{d n}\right)_{n=n_1}.
\end{equation}
 To simplify the calculations, we set $e(0)=0$. Minimizing $F_{GP}$, we find
 \begin{equation}\label{eq:min-ggpe}
   \frac{\delta F_{GP}}{\delta \phi^*(x)} = -\frac{\hbar^2}{2m} \frac{d^2 \phi}{dx^2} + [e'(|\phi|^2) -e'(n_1)] \phi =0,
 \end{equation}
that is integrated into
\begin{equation}
  \label{eq:first-integral-ggpe}
  -\frac{\hbar^2}{2m} \left(\frac{d\phi}{dx}\right)^2 + e(\phi^2) -e'(n_1)\phi^2 = C
\end{equation}
For $x \to -\infty$, $\phi \to 0$ and $d\phi/dx \to 0$, implying $C=0$. That condition also permits $\phi^2(x\to +\infty)=n_1$. By the arguments of App.~\ref{sec:maxwell}, for $0\le \phi^2 \le n_1$, $e(\phi^2) -e'(n_1)\phi^2\ge 0$, and we rewrite~(\ref{eq:first-integral-ggpe})
\begin{equation}
  \label{eq:soliton}
  \int_0^{\phi(x)} \frac{d\phi}{\sqrt{\frac{2m}{\hbar^2} \left[e(\phi^2)-\frac{e(n_1)}{n_1} \phi^2\right]}} = \pm (x-x_0).
\end{equation}
Eq.~(\ref{eq:soliton}) describes a soliton interpolating between $\phi=0$ at $-\infty$ and $\phi=\sqrt{n_1}$ at $+\infty$. The sudden jump of the density given by the Maxwell construction, is replaced by transition region whose width is controlled by quantum fluctuations. In the vicinity of $\phi=0$,
\begin{equation}
  \phi(x \to -\infty) \sim \exp\left[\frac{\sqrt{2m \left(\frac{-e(n_1)}{n_1}\right)}}{\hbar} (x-x_0)\right],
\end{equation}
while, for $\phi^2$ close to $n_1$, $e(\phi^2)-e(n_1) \phi^2/n_1 = e"(n_1)(\phi^2-n_1)^2 + o(\phi^2-n_1)$, leading to
\begin{equation}
  \phi(x\to +\infty)=\sqrt{n_1} \tanh\left[\frac{\sqrt{m e"(n_1) n_1}}{\hbar} (x-x_0)\right].
\end{equation}
The energy associated with the formation of the soliton is
\begin{equation}
  \label{eq:wall-energy}
  E_s=\sqrt{\frac{2\hbar^2}{m}} \int_0^{\sqrt{n_1}} \sqrt{e(\phi^2)-\frac{e(n_1)}{n_1} \phi^2},
\end{equation}
it is finite, due to the exponential decay at $\pm \infty$, and represents the surface tension between the vacuum and the droplet.
When $0> \mu > e'(n_1)$, the profile is given by the equation
 \begin{equation}
  \label{eq:breather}
  \int \frac{d\phi}{\sqrt{\frac{2m}{\hbar^2} \left[e(\phi^2)-\mu \phi^2\right]}} = (x-x_0).
\end{equation}
The amplitude $\phi(x)$ first increases as
\begin{eqnarray}
  \phi(x \to -\infty)= \exp\left[ \frac{2 m |\mu|}{\hbar} (x-x_0)\right],
\end{eqnarray}
until it reaches a maximum for $\phi(x)=\phi_c$ with $e(\phi_c^2)=\mu\phi_c^2$, then it decreases and returns to $\phi=0$.

For $\phi$ close to $\phi_c$, we can approximate
\begin{eqnarray}
  \frac{\hbar}{\sqrt{2m(\mu -e'_0(\phi^2_c))}} \int \frac{d\phi}{\sqrt{\phi_c^2 -\phi^2}} =\pm (x-x_0),
\end{eqnarray}
and find
\begin{eqnarray}
  \phi(x)= \phi_c \cos \left(\frac{\sqrt{2m(\mu -e'(\phi^2_c))}}{\hbar} (x-x_0)\right).
\end{eqnarray}
When $e(\phi_c^2)/\phi_c^2-e'(\phi^2_c)$ is a small quantity, $\phi$ varies slowly in the vicinity of its maximum, giving rise to a near plateau of the density.  This corresponds precisely to a droplet of finite length.  When the size of the droplet is large, it can be treated as a soliton and an antisoliton weakly bound together, with a binding energy decreasing exponentially with the width of the droplet. With larger values of  $e(\phi_c)^2/\phi_c^2-e'_0(\psi^2_c)$, instead of a plateau, a density peak is obtained.

\end{document}